\documentclass[
    ,final          
  ]
  {aipproc}

\layoutstyle{6x9}

\usepackage{amsmath}
\usepackage{amssymb}
\usepackage{amsfonts}
\usepackage{fleqn}

\newcommand{\CenterEps}[2][1]{\ensuremath{\vcenter{\hbox{\includegraphics[scale=#1]{#2.eps}}}}}

\def\<{\left\langle}
\def\>{\right\rangle}
\def\SnI{N}

\newcommand{\bi}{\begin{itemize}}
\newcommand{\ei}{\end{itemize}}

\newcommand{\be}{\begin{equation}}
\newcommand{\ee}{\end{equation}}
\newcommand{\bea}{\begin{eqnarray}}
\newcommand{\eea}{\end{eqnarray}}

\def\<{\left\langle}
\def\>{\right\rangle}

\newcommand{\SuperField}[1]{\hat{#1}}

\allowdisplaybreaks[2]

\begin{document}

\title{Sneutrino Hybrid Inflation}

\author{Stefan Antusch}{ 
address={
Departamento de F\'{\i }sica Te\'{o}rica C-XI 
and 
Instituto de F\'{\i }sica Te\'{o}rica C-XVI, \\
Universidad Aut\'{o}noma de Madrid, 
Cantoblanco, E-28049 Madrid, Spain
}
}

\begin{abstract} 
We review the scenario of sneutrino hybrid inflation, where one of the
singlet sneutrinos, the superpartners of the right-handed neutrinos, 
plays the role of the inflaton. 
In a minimal model of sneutrino hybrid inflation, the spectral index is given by $n_\mathrm{s} \approx 1 + 2 \gamma$. 
With $\gamma = 0.025 \pm 0.01$ constrained by WMAP, a running spectral index 
$|\mathrm{d} n_\mathrm{s}/ \mathrm{d} \ln k| \ll |\gamma| $ 
and a tensor-to-scalar ratio $r \ll \gamma^2$ are predicted. 
Small neutrino masses arise from the seesaw mechanism,
with heavy masses for the singlet (s)neutrinos generated by the vacuum expectation value  of the waterfall field after inflation.  
The baryon asymmetry of the universe can be explained by non-thermal leptogenesis via sneutrino inflaton decay, 
with low reheat temperature $T_\mathrm{RH} \approx 10^{6}$ GeV. 
\end{abstract}

\maketitle

\section{Introduction}

The interface between early universe cosmology and particle physics 
provides many challenges.
Open fundamental questions in this context include the identity of the scalar field responsible for inflation \cite{Guth:1980zm}, 
in order to solve flatness and horizon problems of the early universe, 
and the origin of the observed baryon asymmetry.

The experimental discovery of neutrino mass and mixing, 
when combined with the ideas of 
the see-saw mechanism and supersymmetry, gives a new
perspective on both of these challenges.
In order to generate the observed neutrino masses within 
a see-saw extended version of the Minimal Supersymmetric Standard Model
(MSSM), right-handed neutrinos (together with their scalar superpartners, the singlet sneutrinos) are typically introduced and small neutrino
masses arise naturally from the see-saw mechanism.  
In see-saw scenarios, the out-of-equilibrium decay of these right-handed (s)neutrinos in the early universe can generate the observed baryon asymmetry \cite{Fukugita:1986hr}. 
Among the particles of this extended MSSM, the singlet sneutrinos become 
attractive candidates for playing the role of the inflaton.
Motivated by such considerations, the possibility of chaotic (large
field) inflation with a sneutrino inflaton has been proposed \cite{Murayama:1992ua}. 
An alternative to chaotic inflation is hybrid inflation \cite{Linde:1993cn}, 
which, in contrast to chaotic inflation, involves field 
values well below the Planck scale and is thereby promising for connecting inflation to particle physics.

In this talk we review the scenario proposed in \cite{Antusch:2004hd},  
that one of the 
singlet sneutrinos $\tilde{\SnI}_i$, where $i=1,2,3$ is a family index, 
plays the role of the inflaton field of hybrid inflation. 
We present a minimal model of sneutrino hybrid inflation, investigate prospects of generating the baryon asymmetry of our universe via non-thermal leptogenesis \cite{Fukugita:1986hr,nonthermalLG} and  
discuss how future observations can distinguish sneutrino hybrid inflation from scenarios of chaotic sneutrino inflation.

\section{Superpotential for Sneutrino Hybrid Inflation} 
In order to illustrate how sneutrino hybrid inflation can be realized, 
let us consider the following minimal superpotential
 \begin{eqnarray}\label{eq:W4}
\mathcal{W} = 
\kappa \SuperField{S} \left(\frac{\SuperField{\phi}^4}{{M'}^2} -
M^2\right) \!+
\frac{(\lambda_{N})_{ij}}{M_*}
\,\SuperField{\SnI}_{i}\SuperField{\SnI}_{j}\,\SuperField{\phi}\SuperField{\phi} \, 
+ \dots \;.
\end{eqnarray}
$\kappa$ and $(\lambda_{N})_{ij}$ are dimensionless Yukawa
couplings and $M,{M'}$ and $M_*$ are, in the most general case, 
three independent mass scales. 
The superfields $\SuperField{\SnI}_i$, $\SuperField{\phi}$ and 
$\SuperField{S}$ contain the 
following bosonic components, respectively: 
the singlet sneutrino inflaton $\tilde{\SnI}$ (dropping the family index here and in the following),
which is non-zero during inflation;  
the so-called waterfall 
field $\phi$, which is held at zero during inflation
but which develops a non-zero 
vacuum expectation value (vev) after inflation; 
and the singlet field $S$ which is held practically at zero during and after inflation.

The form of $\mathcal{W}$ in Eq.~(\ref{eq:W4}) can be understood
as follows:
\begin{itemize}
\item The first term on the right-hand side of Eq.~(\ref{eq:W4}) serves to fix 
the vev of the waterfall field after inflation and contributes a large
vacuum energy to the potential during inflation.

\item The second term on the right-hand side of Eq.~(\ref{eq:W4})
allows the sneutrino inflaton to give a positive mass 
squared for the waterfall field during inflation, 
which fixes its vev at zero as long as 
$|\tilde{\SnI}|$ is above a critical value. After inflation, when the 
waterfall field acquires its non-zero vev, the same term
yields the masses of the singlet (s)neutrinos, 
which are required for explaining the smallness of neutrino masses via 
the see-saw mechanism.  
 
\end{itemize}
In order to obtain this minimal form of the superpotential, 
we have chosen the waterfall superfield to appear in this term as
$\SuperField{\phi}^4/{M'}^2$ (instead of 
e.g.\ $\SuperField{\phi}^2$) in this example. This allows to impose  
a $Z_4$ discrete symmetry which prevents explicit singlet (s)neutrino
masses. We would like to note at this point that the dots in Eq.~(\ref{eq:W4}) 
include $Z_4$-violating higher-dimensional operators such as, e.g., 
$\SuperField{S} \SuperField{\phi}^5/M'^3$ 
(or even $\SuperField{S} \SuperField{\phi}^5/m_\mathrm{P}^3$)  
which lift the degeneracy of the true vacuum and effectively blow away 
potential domain wall networks associated with $Z_4$-breaking after inflation.
$\mathcal{W}$ is also compatible with a $U(1)_{\mathrm{R}}$-symmetry under
which $\mathcal{W}$ and $\SuperField{S}$ each carry unit R-charge,
while the charge of $\SuperField{\SnI}$ is $1/2$. Under suitable
conditions the discrete subgroup of this symmetry acts as matter
parity.

\section{The K{\"a}hler Potential}
Since the field values of the inflaton are well below the reduced 
Planck scale $m_\mathrm{P} =1/\sqrt{8\pi G_\mathrm{N}}$, 
we can consider an expansion in powers of $1/m_\mathrm{P}^2$:
\begin{eqnarray}\label{eq:K1}
\lefteqn{ \mathcal{K}  = 
|\SuperField{S}|^2 \!+ |\SuperField{\phi}|^2 \!+ |\SuperField{\SnI}|^2 
\!+\kappa_S \frac{|\SuperField{S}|^4}{4 m_\mathrm{P}^2} 
\!+\kappa_N \frac{|\SuperField{\SnI}|^4}{4 m_\mathrm{P}^2} 
\!+\kappa_\phi \frac{|\SuperField{\phi}|^4}{4 m_\mathrm{P}^2} 
}
\nonumber \\
& \, +& \!\!\kappa_{S\phi} \frac{|\SuperField{S}|^2 |\SuperField{\phi}|^2}{m^2_\mathrm{P}}
\!+ \kappa_{S N} \frac{|\SuperField{S}|^2 |\SuperField{\SnI}|^2}{m^2_\mathrm{P}}
\!+\kappa_{N \phi} \frac{|\SuperField{\SnI}|^2 |\SuperField{\phi}|^2}{m^2_\mathrm{P}}
\!+ \dots \, ,
\end{eqnarray}
where the dots indicate higher order terms and
additional terms for the other fields.  

With non-zero F-terms during inflation, the non-canonical K{\"a}hler potential can 
contribute significantly to the scalar potential.

\section{The Scalar Potential}
Within the model defined by the 
superpotential $\mathcal{W}$ of Eq.\ (\ref{eq:W4}) and the K{\"a}hler 
potential $\mathcal{K}$ of Eq.\ (\ref{eq:K1}), we can now analyze the scalar potential.
The F-term contributions to  
the scalar potential are given by:\footnote{We will neglect radiative
corrections to the potential in the following, which are generically  
subdominant in our model.}  
\begin{eqnarray}
V_\mathrm{F} \;=\; e^{\mathcal{K}/m^2_\mathrm{P}} \left[ 
K_{ij}^{-1} D_{z_i} \mathcal{W} D_{z^*_j} \mathcal{W}^* 
- 3 m_\mathrm{P}^{-2} |\mathcal{W}|^2 
\right] , 
\end{eqnarray}
with $z_i$ being the bosonic components of the superfields
$\SuperField{z}_i \in 
\{\SuperField{\SnI},\SuperField{\phi},\SuperField{S},\dots\}$   
 and where we have 
replaced the superfields in $\mathcal{W}$ 
and $\mathcal{K}$ by their bosonic components and 
 defined   
\begin{eqnarray}
D_{z_i} \mathcal{W} := \frac{\partial  \mathcal{W}}{\partial z_i} + 
 m_\mathrm{P}^{-2} \frac{\partial  \mathcal{K}}{\partial z_i} \mathcal{W} 
 \: , \; K_{ij} := \frac{\partial^2 \mathcal{K}}{\partial z_i \partial z^*_j} 
\end{eqnarray}
and $D_{z^*_j} \mathcal{W}^* := (D_{z_j} \mathcal{W})^*$.
Since we assume that 
 $\tilde{\SnI}, \phi$
 and $S$ are effective gauge singlets at the energy scales under consideration, 
 there are no relevant D-term contributions.   
 From Eqs.\ (\ref{eq:W4}) and (\ref{eq:K1}), 
 with canonically normalized fields, and writing the potential in terms of real fields 
$\tilde{\SnI}_\mathrm{R}=\sqrt{2}|\tilde{\SnI}|$, 
$\phi_\mathrm{R}= \sqrt{2} |\phi|$ and $S_\mathrm{R}=\sqrt{2}|S|$, 
 we obtain
\begin{eqnarray}\label{eq:scalarpot2}
V &=& \kappa^2 \left(\frac{\phi_\mathrm{R}^4}{4 {M'}^2} - M^2 \right)^2 
\left(
1 -\beta \frac{ \phi_\mathrm{R}^2}{2 m_\mathrm{P}^2} 
+\gamma \frac{\tilde{\SnI}_\mathrm{R}^2}{2 m_\mathrm{P}^2} 
- \kappa_S \frac{ S_\mathrm{R}^2 }{2 m_\mathrm{P}^2}  \right)
 \nonumber \\
&&
 + \frac{\lambda_N^2}{2 M_*^2} \,(\tilde{\SnI}_\mathrm{R}^4 \phi_\mathrm{R}^2 
+ \tilde{\SnI}_\mathrm{R}^2 \phi_\mathrm{R}^4) + \dots  
\,, 
\end{eqnarray} 
where we have defined 
\begin{eqnarray}
\label{eq:beta} \beta &:=&\kappa_{S\phi}-1 \quad \mbox{($>0$ for inflation to
end)}\; ,\\
\gamma &:=& 1-\kappa_{S N}\;.
\end{eqnarray} 

\section{Realizing Sneutrino Hybrid Inflation}
From Eq.~(\ref{eq:scalarpot2}) we see that $S_\mathrm{R}$ can be set to zero 
during inflation 
if we take, e.g.,  $\kappa_S < -1/3$, such that $S_\mathrm{R}$ gets a mass term 
larger than the Hubble parameter $H \approx \sqrt{V_0}/(\sqrt{3}m_\mathrm{P})$. 
With $\phi_\mathrm{R} = S_\mathrm{R} = 0$, the part of
the scalar potential relevant for the evolution of the singlet sneutrino 
inflaton $\tilde{\SnI}_\mathrm{R}$ during inflation is given by
\begin{eqnarray}\label{eq:scalarpot3}
V &=& \kappa^2 M^4 \left(
1 
+\gamma \frac{\tilde{\SnI}_\mathrm{R}^2}{2 m_\mathrm{P}^2} 
+ \delta \frac{\tilde{\SnI}_\mathrm{R}^4}{4 m_\mathrm{P}^4}\right) + \dots \;  , 
\end{eqnarray}
where we have included the next-to-leading order term proportional to 
$\delta$.

\begin{figure}
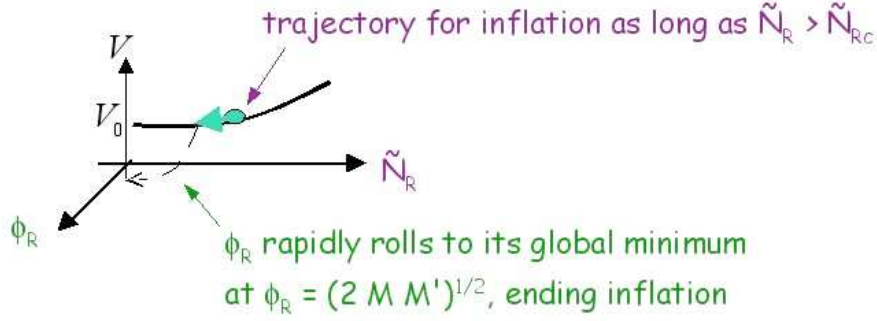
 
\CenterEps[0.6]{endofinflation}
  \caption{\label{fig:endofinflation}
Inflationary trajectory and end of inflation by a second order phase transition
in sneutrino hybrid inflation. 
 }
\end{figure}

During inflation, the waterfall field $ \phi_\mathrm{R}$ has a zero vev and the
potential is dominated by the vacuum energy $V_0 = \kappa^2 M^4$. 
This false vacuum during inflation is stable as long as the mass squared for the waterfall
field $\phi_\mathrm{R}$ is positive. From Eq.~(\ref{eq:scalarpot2}) we obtain
 the requirement  
\begin{eqnarray}
m_{\phi_\mathrm{R}}^2 = 
\lambda_N^2 \frac{\tilde{\SnI}_\mathrm{R}^4}{M_*^2} - 
\beta \frac{\kappa^2 M^4 }{m_\mathrm{P}^2} \;>\; 0\; .
\end{eqnarray}
Inflation thus ends when 
the squared mass of the waterfall field becomes negative, i.e.\ 
$\phi_\mathrm{R}$ develops a
tachyonic instability and rolls rapidly to its global minimum at 
$\<\phi_\mathrm{R}\> = \sqrt{2 {M'} M}$ (as illustrated in 
Fig.~\ref{fig:endofinflation}). 
Clearly, this requires $\beta > 0$, as already indicated in
Eq.~(\ref{eq:beta}). 
More precisely, inflation ends by a second
order phase transition when the field value of the inflaton drops 
below the critical value $\tilde{\SnI}_{\mathrm{R} c}$ given by 
\begin{equation} \label{eq:nucrit}
\tilde{\SnI}_{\mathrm{R} c}^2=  \sqrt{\beta}\frac{\kappa}{\lambda_N}\frac{
M^2  M_*}{m_\mathrm{P}} \;.  
\label{phinuc}
\end{equation}
The field value of the inflaton $\tilde{\SnI}_\mathrm{R}$ at 
$N=$ 50 to 70 $e$-folds before the end of inflation is then given approximately by  
\begin{equation}\label{eq:Nuat60Ne}
\tilde{\SnI}_{\mathrm{R}e} \;\approx\; \tilde{\SnI}_{\mathrm{R} c} 
\,e^{\gamma N }\,.
\label{phinue}
\end{equation}

During inflation, the parameter $\gamma$ in the scalar potential in Eq.~(\ref{eq:scalarpot3}) controls the mass of the 
inflaton. 
Furthermore, compared to the term proportional to $\gamma$, the term proportional
to $\delta$ is suppressed by $\tilde{\SnI}_\mathrm{R}^2/m_\mathrm{P}^2$. 
The slow-roll parameters are 
given by  
\begin{eqnarray}
\!\!\!\!\!\! \epsilon \!&:=&\!  
 \frac{m_\mathrm{P}^2}{2} \!\left(\frac{V'}{V}\right)^2
\!\!\!\simeq \frac{(\delta \tilde{\SnI}_\mathrm{R}^3 \!+ m_\mathrm{P}^2 \gamma  
\tilde{\SnI}_\mathrm{R})^2}{2 m_\mathrm{P}^6}
\approx \gamma^2 \!\frac{\tilde{\SnI}_\mathrm{R}^2}{2 m_\mathrm{P}^2} , \\
\!\!\!\!\!\!\eta \!  &:= &\!   m_\mathrm{P}^2\!\left(\frac{V''}{V}\right) 
\simeq \gamma + \frac{3 \,\delta \tilde{\SnI}_\mathrm{R}^2}{m_\mathrm{P}^2}
\approx \gamma \, , \\
\!\!\!\!\!\!\xi \!  &:=&\!   m_\mathrm{P}^4 \!\left(\frac{V'\, V'''}{V^2}\right)
 \simeq \frac{6\, \delta \tilde{\SnI}_\mathrm{R}^2 (\gamma m_ \mathrm{P}^2 + 
 \delta \tilde{\SnI}_\mathrm{R}^2)}{m_\mathrm{P}^4} ,
\end{eqnarray}
where prime denotes derivative with respect to $N_R$.  
Thus, assuming that the slow-roll approximation  
is justified (i.e.\ $\epsilon \ll 1$, $\eta \ll 1$), the spectral index 
$n_\mathrm{s}$, the tensor-to-scalar ratio
$r=A_\mathrm{t}/A_\mathrm{s}$ and the running spectral index 
$\mathrm{d} n_\mathrm{s}/\mathrm{d} \ln k$ are given by 
\begin{eqnarray}
n_\mathrm{s} \!&\simeq&\! 1 - 6 \epsilon + 2 \eta \;\approx\; 1 + 2 \gamma\; ,
\vphantom{\frac{1}{1}}\\
r \!&\simeq&\! 16 \epsilon \approx  \gamma^2 \, 
\frac{8 \,\tilde{\SnI}_{\mathrm{R}e}^2}{ m_\mathrm{P}^2}\; , \\
\frac{\mathrm{d} n_\mathrm{s}}{\mathrm{d} \ln k} \!\!&\simeq&\!\! 
 16 \epsilon \eta - 24 \epsilon^2 - 2 \xi 
\approx 
-\gamma \, \frac{12  \delta\,\tilde{\SnI}_{\mathrm{R}e}^2}{ m_\mathrm{P}^2}
\; .
\end{eqnarray}

\section{Constraints from Experimental Data}
\begin{figure}
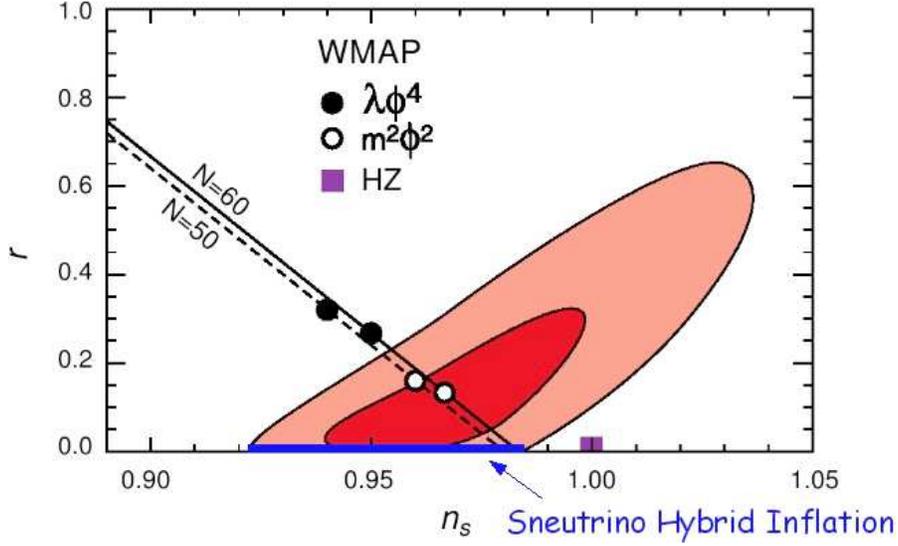
 
\CenterEps[0.6]{tensortoscalar}
  \caption{\label{fig:tensortoscalar}
Graphical illustration of the predictions for the tensor-to-scalar ratio $r$ and the 
spectral index $n_\mathrm{s}$ in sneutrino hybrid inflation \cite{Antusch:2004hd} (narrow strip on the bottom) and in chaotic sneutrino inflation \cite{Murayama:1992ua} with a quadratic potential (white dots),  
compared to the allowed regions by WMAP \cite{Spergel:2006hy}. 
In sneutrino hybrid inflation, the tensor-to-scalar ratio $r$ is predicted to be $r \ll \gamma^2$, with $\gamma = 0.025 \pm 0.01$ from the experimental data on the spectral index, as discussed in the text. By their predictions for $r$, future observations will be able to distinguish sneutrino hybrid inflation from chaotic sneutrino inflation.
 }
\end{figure}

The experimental data on the spectral index from WMAP 
$n_\mathrm{s}=0.95 \pm 0.02$ \cite{Spergel:2006hy} restricts 
$\gamma$ to be roughly $\gamma = 0.025 \pm 0.01$. 
As discussed above, $\gamma$ controls the sneutrino mass during inflation. 
In this model it stems mainly from
supergravity corrections. 
In addition, we see that
the tensor-to-scalar ratio
$r=A_\mathrm{t}/A_\mathrm{s}$ and the running spectral index 
$\mathrm{d} n_\mathrm{s}/\mathrm{d} \ln k$ are suppressed by higher powers of
$\gamma$ or by $\tilde{\SnI}_\mathrm{R}^2/m_\mathrm{P}^2$ and are thus generically small.
Especially the prediction for the tensor-to-scalar ratio $r \ll \gamma^2$ 
is thus in sharp contrast to the prediction of $r \approx 0.16$ for the case of 
chaotic sneutrino inflation with a quadratic superpotential, as illustrated in Fig.~\ref{fig:tensortoscalar}. 

In our model, the amplitude of the 
primordial spectrum is given by 
\begin{equation}
P_{\cal R}^{1/2} \simeq 
\frac{1}{\sqrt{2 \varepsilon}} \left(\frac{H}{2\pi
m_{\mathrm{P}}}\right)
\approx \frac{\kappa}{2 \sqrt{3} \,\gamma \,\pi}
\frac{M^2}{m_\mathrm{P}\, \tilde{\SnI}_{\mathrm{R}e}} \,.
\label{spectrum}
\end{equation} 
 Given the COBE normalization $P_{\cal R}^{1/2}\approx 5 \times 10^{-5}$ 
 \cite{Smoot:1992td},
from Eqs.\ (\ref{phinuc}), 
 (\ref{spectrum}) and (\ref{phinue}) we obtain 
\begin{equation}\label{eq:scales}
\frac{M^2}{M_*\,m_\mathrm{P}} \approx \:3 \times 10^{-8} \,
\frac{\gamma^2\,\sqrt{\beta}}{\kappa \, \lambda_N}\;,
\end{equation}
which relates the scale $M$ in the superpotential to the cutoff scale $M_*$. 
It has to be combined with the constraint 
$\tilde{\SnI}_{\mathrm{R}e}\ll m_\mathrm{P}$
 (see Eqs.~(\ref{eq:Nuat60Ne}) and (\ref{eq:nucrit})) and with 
 $M < {M'}, M_*$.

\section{Reheating and Non-thermal Leptogenesis}
An interesting feature of sneutrino inflation  
is that the observed baryon asymmetry can arise via non-thermal leptogenesis \cite{Fukugita:1986hr,nonthermalLG} directly through sneutrino inflaton decay \cite{SIleptogenesis}.
To illustrate the mechanism, let us assume in the following the situation that the inflaton is the lightest
singlet sneutrino $\tilde{\SnI}_1$ and that it dominates leptogenesis and reheating 
after inflation. 
In sneutrino hybrid inflation \cite{Antusch:2004hd}, this is, e.g., the case if the waterfall field $\phi$ decays earlier than the singlet sneutrino inflaton via heavier 
singlet neutrinos $\SnI_2$ (or $\SnI_3$) with comparably 
large couplings to $\phi$. 

From Eq.~(\ref{eq:W4}), using $\<\phi \> = \sqrt{{M'} M}$, we see
that the mass of $\tilde{\SnI}_1$ is given by $M_\mathrm{R1} = 2 (\lambda_N)_{11} {M'}M/M_*$ in
the basis where the mass matrix $M_\mathrm{R}$ of the singlet (s)neutrinos is 
diagonal. 
It decays mainly via the extended MSSM
Yukawa coupling $(Y_\nu)_{i1}
\SuperField{L}_{i} \SuperField{H}_\mathrm{u} \SuperField{\SnI}_1$ into
slepton and Higgs or into lepton and Higgsino with a decay width given
by $\Gamma_{\SnI_1} = M_\mathrm{R1} {(Y^\dagger_\nu Y_\nu)_{11}}/{(4\pi)}$. 
The decay of the singlet sneutrino after inflation reheats the universe to a
temperature $T_{\mathrm{RH}} \approx (90/(228.75\pi^2 ))^{1/4} \sqrt{
\Gamma_{\SnI_1} m^{}_{\mathrm{P}}}$. 
If $M_\mathrm{R1}\gg T_{\mathrm{RH}}$, the lepton asymmetry is produced
via cold decays of the singlet sneutrinos and the produced baryon asymmetry can be estimated as
${n_\mathrm{B}}/{n_\gamma} \approx - 1.84 \, \varepsilon_1\,
{T_{\mathrm{RH}}}/{M_\mathrm{R1}}$, where $\varepsilon_1$ is the decay
asymmetry for the singlet sneutrino $\tilde{\SnI}_1$.

To take a concrete example of the above discussion, neutrino Yukawa
couplings $(Y_\nu)_{i1}\approx 10^{-6}$ and a sneutrino mass of
$M_\mathrm{R1}=10^8$ GeV 
allow to generate the observed baryon asymmetry of the universe  
$n_\mathrm{B} /n_\gamma 
\approx (6.1\,\pm\,0.2)\,\times\,10^{-10}$ \cite{Spergel:2006hy} and 
imply a reheat temperature
$T_{\mathrm{RH}} \approx 10^6$ GeV. Such a low reheat temperature  
is desirable with respect to gravitino constraints 
(see e.g.\ \cite{gravitinoproblem}) in some supergravity models.

\section{Summary and Conclusions}
We have reviewed the scenario of sneutrino hybrid inflation, where the
singlet sneutrino, the superpartner of the right-handed 
neutrino, plays the role of the inflaton. Sneutrinos are present in any extension of the MSSM, where the smallness of the observed neutrino masses is explained via 
the see-saw mechanism.  
In a minimal model of sneutrino hybrid inflation in  
supergravity, we have found a spectral index $n_\mathrm{s} \approx 1 + 2 \gamma$ 
with $\gamma = 0.025 \pm 0.01$ constrained by WMAP, leading to the prediction   
$|\mathrm{d} n_\mathrm{s}/ \mathrm{d} \ln k| \ll |\gamma| $
for the running spectral index and  $r \ll \gamma^2$ for the 
tensor-to-scalar ratio. 
The prediction for the tensor-to-scalar ratio in sneutrino hybrid inflation 
is thus much smaller than the prediction $r \approx 0.16$ of chaotic sneutrino
 inflation and makes sneutrino hybrid inflation   
 easily distinguishable from chaotic sneutrino inflation by future observations. 
In contrast to
chaotic inflation, the field values of the singlet sneutrino inflaton in hybrid
inflation are well
below the Planck scale. 
We have discussed how the baryon asymmetry of our universe can be 
explained via non-thermal leptogenesis and how a low reheat temperature 
$T_\mathrm{RH} \approx 10^{6}$ GeV can be realized with neutrino Yukawa 
couplings consistent with first family quark and lepton Yukawa couplings in 
Grand Unified Theories.

\begin{theacknowledgments}
I would like to thank  Mar Bastero-Gil, Steve F.~King and Qaisar Shafi
for the collaboration in the work presented here. 
I would also like to thank the organizers of 
DSU 2006 and acknowledge supported by the EU 6$^\text{th}$
Framework Program MRTN-CT-2004-503369 ``The Quest for Unification:
Theory Confronts Experiment''.
\end{theacknowledgments}

\end{document}